\documentclass[aps,prx,twocolumn, superscriptaddress]{revtex4-2}

\usepackage{cmap}
\usepackage{graphicx}
\usepackage{amssymb}
\usepackage{amsmath}
\usepackage{color}
\usepackage{xcolor}
\usepackage{graphicx}
\usepackage{dcolumn}
\usepackage{bm}
\usepackage{natbib}
\usepackage[colorlinks=true,citecolor=blue]{hyperref}
\usepackage{siunitx}

\begin{document}

\title{Coherence of an Electronic Two-Level System under Continuous Charge Sensing by a Quantum Dot Detector}

\author{Subhomoy Haldar}
\email{subhomoy.haldar@ftf.lth.se}
\address{NanoLund and Solid State Physics, Lund University, Box 118, 22100 Lund, Sweden}
\author{Morten Munk}
\address{Physics Department and NanoLund, Lund Universityd, Box 118, 22100 Lund, Sweden}
\author{Harald Havir}
\address{NanoLund and Solid State Physics, Lund University, Box 118, 22100 Lund, Sweden}
\author{Waqar Khan}
\address{NanoLund and Solid State Physics, Lund University, Box 118, 22100 Lund, Sweden}
\address{Presently at Low Noise Factory AB, 41263 Göteborg, Sweden}
\author{Sebastian Lehmann}
\address{NanoLund and Solid State Physics, Lund University, Box 118, 22100 Lund, Sweden}
\author{Claes Thelander}
\address{NanoLund and Solid State Physics, Lund University, Box 118, 22100 Lund, Sweden}
\author{Kimberly A. Dick}
\address{NanoLund and Solid State Physics, Lund University, Box 118, 22100 Lund, Sweden}
\address{Center for Analysis and Synthesis, Lund University, Box 124, 22100 Lund, Sweden}
\author{Peter Samuelsson}
\address{Physics Department and NanoLund, Lund Universityd, Box 118, 22100 Lund, Sweden}
\author{Patrick P. Potts}
\address{Department of Physics and Swiss Nanoscience Institute, University of Basel,
Klingelbergstrasse 82, 4056 Basel, Switzerland}
\author{Ville F. Maisi}
\email{ville.maisi@ftf.lth.se}
\address{NanoLund and Solid State Physics, Lund University, Box 118, 22100 Lund, Sweden}
\date{\today}

\begin{abstract}
We investigate experimentally the quantum coherence of an electronic two-level system in a double quantum dot under continuous charge detection. The charge-state of the two-level system is monitored by a capacitively coupled single quantum dot detector that imposes a back-action effect to the system. The measured back-action is well described by an additional decoherence rate, approximately linearly proportional to the detector electron tunneling rate. We provide a model for the decoherence rate arising due to level detuning fluctuations induced by detector charge fluctuations. The theory predicts a factor of two lower decoherence rate than observed in the experiment, suggesting the need for a more elaborate theory accounting for additional sources of decoherence. 
\end{abstract}

\maketitle

Continuous, weak measurements of quantum systems \cite{wiseman_milburn_2009} allow for investigations of fundamental concepts such as quantum trajectories \cite{murch2013} and quantum state reconstruction \cite{PhysRevA.93.012109}. Combined with feedback, continuous measurements also have a key role in quantum technologies, e.g. providing means for quantum error correction \cite{Livingston2022} and quantum metrology \cite{PhysRevLett.125.200505}. However, the constant coupling of a quantum system to a measurement apparatus, or detector, puts strong requirements on  both detector and coupling properties in order to minimize unwanted back-action effects. This is particularly important in solid state quantum systems, such as superconducting \cite{squbitreview} or semiconductor spin \cite{RevModPhys.95.025003} or charge \cite{semiqubitrev} quantum bits, where sources of decoherence are abundant and can couple into the system via the detector, suppressing quantum coherence.

The arguably most common solid state realization of a continuous measurement is real-time monitoring of the charge state of the system via a capacitively coupled charge detector. Various charge detectors based on quantum point contacts \cite{PhysRevB.67.161308,QPCdet1,PhysRevB.79.035314} and single electron transistors \cite{SET1,SET2} have been demonstrated. Real time charge detection has been employed to investigate e.g. charge and energy transport statistics \cite{PhysRevLett.96.076605,Fujisawa2006}, quasiparticle dynamics in superconducting systems \cite{Ranni2021,Mannila2022}, and spin-to-charge readout in spin qubits \cite{QPCspincharge}. Experimentally, effects of real-time charge detection on electron transport statistics \cite{e12071721,Li2013} and non-equilibrium properties of quantum dots \cite{PhysRevB.79.035303,IHN2010803} have been investigated. However, despite intense theoretical interest~\cite{PhysRevLett.79.3740,PhysRevB.56.15215,PhysRevB.61.2737,PhysRevB.63.125326,oxtoby2006,PhysRevB.81.075301,hell2014,hell2016} there are to date no experimental investigations of how charge detection affects quantum coherence, leaving the potential for charge detection as a quantum technology tool \cite{degen2017} largely unexplored.      

\begin{figure}[b!b]
\includegraphics[width=0.48\textwidth]{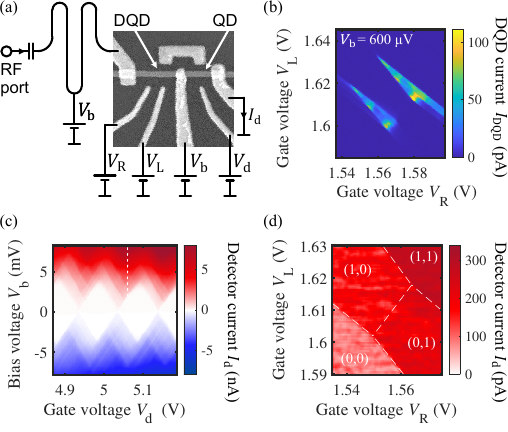}
\caption{\label{fig1} (a) A scanning electron micrograph of the DQD - QD detector device. The DQD hosting a two-level charge qubit system connects to a microwave resonator probed via an RF port on the left. Gate voltages $V_\mathrm{L}$ and $V_\mathrm{R}$ control the occupancies of the DQD and $V_\mathrm{d}$ the detector QD. A bias voltage $V_\mathrm{b}$ is applied to both sides of the DQD to bias the detector but leave the DQD unbiased. The detector current $I_\mathrm{d}$ is measured from the drain of the QD. (b) DQD current $I_\mathrm{DQD}$ as a function of $V_\mathrm{L}$ and $V_\mathrm{R}$. In contrast to the other measurements presented, this one is done such that the current is measured from the DQD side and the QD is kept unbiased to probe the DQD transport. (c) Detector current as a function of $V_b$ and $V_\mathrm{d}$. (d) Detector current $I_\mathrm{d}$ as a function of the DQD gate voltages $V_\mathrm{L}$ and $V_\mathrm{R}$. The detector is tuned to a charge sensitive point with sensitivity $\mathrm{d}I_\mathrm{d}/\mathrm{d}V_\mathrm{d} = -200$~nS by applying $V_\mathrm{b} = 1.4$~mV and tuning $V_\mathrm{d}$ to the steepest point of the Coulomb resonance at $V_\mathrm{d} = \SI{5.05}{V}$.}
\end{figure}

In this work we fill this gap by investigating the coherence of an electronic two-level system in a double quantum dot (DQD), subjected to continuous charge detection via a capacitively coupled single quantum dot (QD). An increase of the DQD decoherence rate captures the changes in the measured response, hence showing that the detector back-action induces decoherence. The decoherence rate increases approximately linearly with the detector current. A microscopic theory including detector back-action provides an expression for the decoherence rate as a function of detector properties, in qualitative agreement with the experimental results but quantitatively underestimating the observed rate by roughly a factor of two. This suggests that the model does not fully capture all decoherence mechanisms. These findings show that our DQD-detector system is a versatile testbed, with tuneable decoherence, for both fundamental experiments on quantum measurement and quantum technology applications based on continuous charge detection. 



The DQD two-level system - QD detector device of Fig.~\ref{fig1}(a) is formed in an InAs polytype nanowire where wurzite segments form tunneling barriers between the zinc blende quantum dots and contacts~\cite{chen2017, barker2019}. The DQD is, on one hand, connected to a QD acting as a charge detector to probe its charge state~\cite{barker2022} and, on the other hand, to a resonator depicted on the left to probe its coherence properties~\cite{khan2021, childress2004, frey2012}.
Figure~\ref{fig1}~(b) shows the electrical current $I_\mathrm{DQD}$ through the DQD as a function of the two plunger gate voltages $V_\mathrm{L}$ and $V_\mathrm{R}$ applied to the gates controlling the occupancy of the two dots. A bias voltage of $V_\mathrm{b} = \SI{600}{ \micro V}$ is applied across the DQD and the current is measured from the resonator side to avoid contributions to the current from the detector QD. Note that for this measurement, the biasing and current measurement differs from the setting of Fig.~\ref{fig1}(a) that was used for all the other measurements. The response of Fig.~\ref{fig1}(b) exhibits triangular regions where transport is allowed. These are the finite bias triangles, the standard feature exemplary of DQD transport.

Figure~\ref{fig1}~(c) on the other hand presents similar transport measurement of the QD detector side. Here the current $I_\mathrm{d}$ is measured on the detector side and the occupancy of the QD is tuned with the detector plunger gate voltage $V_\mathrm{d}$. The current $I_\mathrm{d}$ though the QD forms so-called Coulomb diamonds with suppressed current in the $V_\mathrm{b}$-$V_\mathrm{d}$ plane, characteristic for a single quantum dot. To perform charge detection of the DQD, a fixed bias voltage of $V_\mathrm{b} = 1.4$~mV is applied and the gate voltage is tuned to the falling edge of the finite current regime between two of the Coulomb diamonds, taking place at the gate setting indicated by the dashed line. In this point $I_\mathrm{d}$ is sensitive to changes in $V_\mathrm{d}$ with sensitivity $\mathrm{d}I_\mathrm{d}/\mathrm{d}V_\mathrm{d} = -200$~nS. Therefore the detector also responds to individual electrons moving in the vicinity of the detector. This is shown in Fig.~\ref{fig1}(d) that plots $I_\mathrm{d}$ as a function of the DQD gates at the same operation point as in Fig.~\ref{fig1}(b). We observe a step-wise increase in $I_\mathrm{d}$ whenever an electron is added to the DQD. When adding an excess electron to the left dot, i.e. changing from state $(0,0)$ to $(1,0)$ the current increases by $90$~pA and similarly when adding another to the right dot and moving from state $(1,0)$ to $(1,1)$ by $100$~pA. Similarly, the interdot transition $(1,0) \longleftrightarrow (0,1)$ has a $\Delta I_\mathrm{d} = 34$~pA change in the detector for this operation point. With the lever arm $\alpha_\mathrm{d} = 30\ \mathrm{meV/V}$ determined from the Coulomb diamonds of Fig.~\ref{fig1}(c), the corresponding change in the position of the energy level of the detector is $\Delta E_\mathrm{d} = \alpha_\mathrm{d}\:(\mathrm{d}V_\mathrm{d}/\mathrm{d}I_\mathrm{d})\: \Delta I_\mathrm{d} = \SI{5}{\micro eV}$ for the interdot transition.

\begin{figure}[t]
\includegraphics[width=0.48\textwidth]{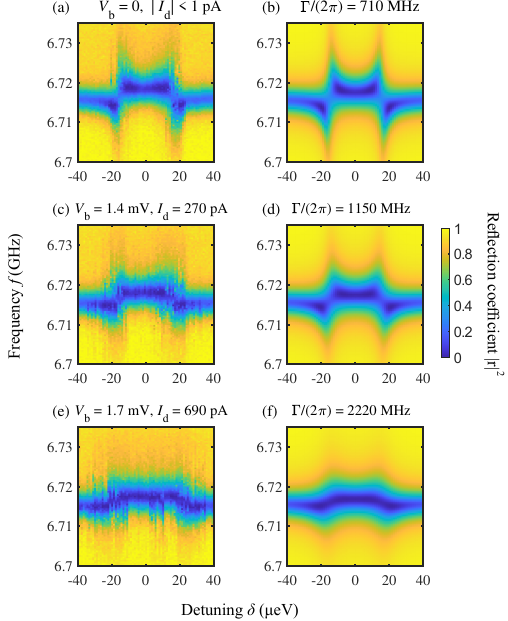}
\caption{\label{fig2} (a) Measured reflection coefficient $|r|^2$ at the resonator input port as a function of drive frequency $f$ and detuning $\delta$. The detector is kept in Coulomb blockade. With the used input power, $P = 10$~aW, the photon number in the cavity is approximately $n = 0.14$ and microwave voltage amplitude $\SI{1.5}{\micro V}$. (b) Calculated response based on Eqs.~(\ref{eq:R}) and (\ref{eq:A}) with the parameter values given in the main text. (c) Resonator response with $V_\mathrm{b} = 1.4$~mV resulting in $I_\mathrm{d} = 270$~pA. d), Corresponding calculation with $\Gamma/{2\pi} = 1150$~MHz and other parameter values as in (b). (e),(f) Same as (c) and (d) but with $V_\mathrm{b} = 1.7$~mV, $I_\mathrm{d} = 690$~pA, and $\Gamma/{2\pi} = 2220$~MHz.}
\end{figure}

Next we turn to measuring the coherence properties of the DQD. The charge states $(1,0)$ and $(0,1)$ define a two-level system. The interdot tunnel coupling $t$ hybridizes these states and gives rise to the energy difference $E = \sqrt{\delta^2 + (2t)^2}$ with energy detuning $\delta$ between the non-hybridized states~\cite{childress2004}. Changing the gate voltage $V_\mathrm{R}$ across the interdot transition line of Fig.~\ref{fig1}(d) changes the detuning as $\delta = \alpha_\mathrm{R}\: \Delta V_\mathrm{R}$, where the lever arm for the DQD plunger gate voltage $V_\mathrm{R}$ is $\alpha_\mathrm{R} = \SI{600}{\micro eV}/\SI{48}{mV} = 12.5\ \mathrm{meV/V}$, determined from Fig.~\ref{fig1}(b). The microwave resonator probes the coherence of this two-level system via capacitive dipole coupling~\cite{khan2021, childress2004, frey2012}. Figure~\ref{fig2}~(a) presents the resulting reflection coefficient measurement as a function of the drive frequency $f$ and detuning $\delta$ with the detector QD kept in Coulomb blockade with vanishing $I_\mathrm{d}$. 

To obtain the DQD parameter values, we fit the measurement of Fig.~\ref{fig2}(a) with input-output theory of the coupled DQD-resonator system with a Jaynes-Cummings Hamiltonian and disregarding the detector dot~\cite{frey2012, khan2021, ranni2023}. In the limit of small drive power, see Ref.~\citealp{ranni2023}, the reflection coefficient takes the simple from
\begin{equation}
    \label{eq:R}
    \left|r\left(\omega\right)\right|^2 = \left| \kappa_c A\left(\omega\right) - 1 \right|^2,
\end{equation}
with
\begin{equation}
    \label{eq:A}
    A\left(\omega\right) = \frac{\Gamma/2 - i\left(\omega - \omega_q \right)}{\left[\kappa/2 - i\left(\omega - \omega_c \right) \right] 
    \left[\Gamma/2 - i\left(\omega - \omega_q \right) \right] + g^2}.
\end{equation}
Here $\omega = 2 \pi f$ is the drive frequency, $\omega_c$ the resonance frequency, $\kappa_c$ the resonator input coupling, and $\kappa = \kappa_c + \kappa_i$ the resonator linewidth with internal resonator losses of $\kappa_i$. The rest of the parameters describe the DQD: $\Gamma$ is the decoherence rate of the DQD, $g$ the dipole coupling coefficient between the resonator and the DQD and $\omega_q = E/\hbar$ the energy difference of the DQD states. Figure~\ref{fig2}~(b) presents the fit of Eq.~(\ref{eq:R}) to the measurement of Fig.~\ref{fig2}(a). First, the bare resonator parameters $\omega_c/2\pi = \SI{6.716}{GHz}$, $\kappa_c/2\pi = \SI{3.6}{MHz}$ and $\kappa_i/2\pi = \SI{1.4}{MHz}$ are determined in the far detuned case at $|\delta| \geq \SI{40}{\micro eV}$. Then, the inter-dot tunnel coupling $t = h \times \SI{2.8}{GHz}$ sets the resonance condition $\omega_q = \omega_c$ at $\delta = \pm \SI{16}{\micro eV}$, the coupling $g/2\pi = \SI{44}{MHz}$ determines the magnitude of the DQD response in $f$ and the decoherence rate $\Gamma/2\pi = \SI{710}{MHz}$ gives its smearing. See also Refs.~\citealp{frey2012, stockklauser2015, scarlino2022} that use similar procedure based on numerical calculations.

\begin{figure}[t]
\includegraphics[width=0.48\textwidth]{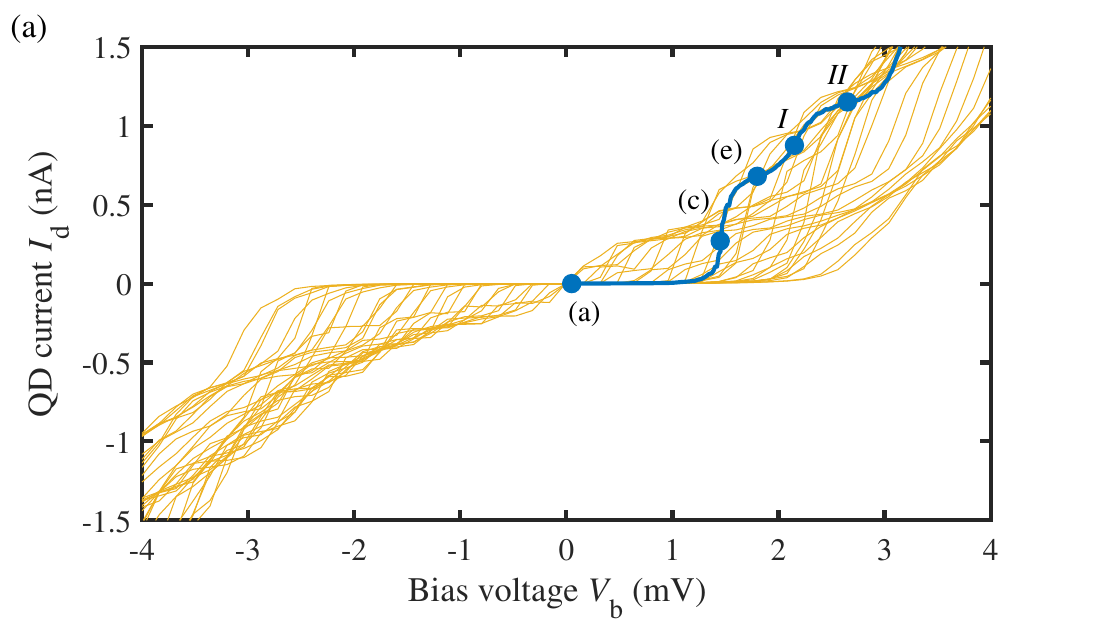}
\includegraphics[width=0.48\textwidth]{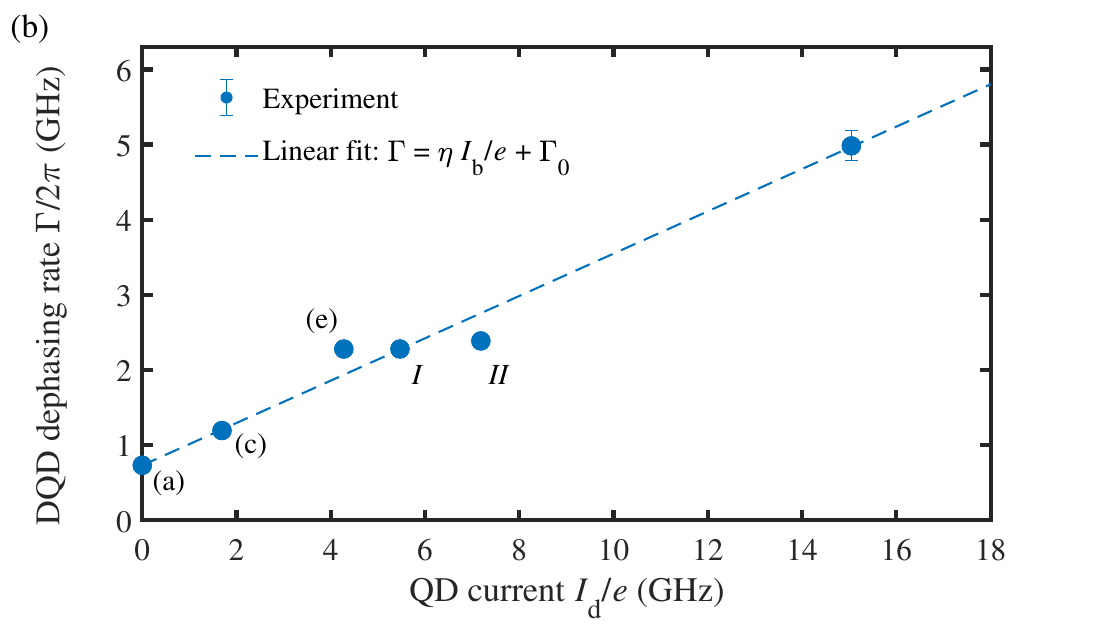}
\caption{\label{fig3} (a), Detector response around the Coulomb resonance at $V_G = 5.1$~V. Each yellow curve shows the envelope of $I_\mathrm{d}$ for a given $V_\mathrm{d}$ in the range $5.008 - 5.074$~V. The blue curve at $V_\mathrm{d} = 5.050$~V highlights the detector operation points used in this study. The blue dots marked with (a) - (c) correspond to the measurements of Fig.~\ref{fig2}. The points marked with {\it I} and {\it II} denote two points further at the second step of the $I_\mathrm{d}-V_\mathrm{b}$ curve. (b) The DQD decoherence rate $\Gamma/2\pi$ as a function of detector QD current $I_\mathrm{d}$. The blue dots are experimental data based on the fitting procedure of Fig.~\ref{fig2} and the dashed blue line a linear fit with slope $\eta = 1.8$ and decoherence rate $\Gamma_0/2\pi = 710$ MHz for $I_\mathrm{d} = 0$.}
\end{figure}

Now we bias the detector and set it to a typical charge sensor operation mode, see Fig.~\ref{fig3}(a): Applying a bias voltage $V_\mathrm{b} = 1.4$~mV of approximately half of the size of the Coulomb diamond and tuning the gate voltage into the edge of the Coulomb diamond at point marked with "(c)" gives rise to the steep slope in $I_\mathrm{d}$ with the sensitivity $\mathrm{d}I_\mathrm{d}/\mathrm{d}V_\mathrm{d} = -\alpha_\mathrm{d}\: \mathrm{d}I_\mathrm{d}/\mathrm{d}V_\mathrm{b} = -200$ nS used also for the measurement of Fig.~\ref{fig1}(d). Figure~\ref{fig2}~(c) shows the effect of this charge detection operation point to the measured DQD response. The fit of Fig.~\ref{fig2}~(d) catches the changes in the response by keeping all the other parameter values fixed except the total decoherence rate with fitted value of $\Gamma/2\pi = 1150$~MHz. The detector adds decoherence to the DQD. Figures~\ref{fig2}~(e) and (f) repeat the experiment with bias voltage increased to $V_b = 1.7$~mV. With this configuration, see Fig.~\ref{fig3}(a), the detector current is approximately doubled and the operation point sits in the flat part between the two steps suppressing the sensitivity to $\mathrm{d}I_\mathrm{d}/\mathrm{d}V_\mathrm{d} = -4$~nS. Despite the suppressed charge sensitivity, the DQD decoherence rate increased to $\Gamma/2\pi = 2220$~MHz. 

Figure~\ref{fig3}~(b) summarizes the above measurements by plotting $\Gamma$ as a function of the detector current $I_\mathrm{d}$. We continued here the measurements to the next rising step ({\it I}) with $\mathrm{d}I_\mathrm{d}/\mathrm{d}V_\mathrm{d} = -19$~nS and plateau ({\it II}) with $\mathrm{d}I_\mathrm{d}/\mathrm{d}V_\mathrm{d} = -4$~nS. Interestingly, though the charge sensitivity as well as the detector current $I_\mathrm{d}$ both increase from point (e) to {\it I}, the decoherence remains constant. Continuing the experiment to even larger detector bias, the increasing trend is recovered at much higher bias of $V_b = 3.9$~mV with $I_d = 2.4$~nA. 

To analyze the effect of the detector backaction on the DQD coherence, we extend the theoretical model behind Eq.~(\ref{eq:R}) to include the detector QD and its charge dynamics, see \cite{SI} for details. Due to the capacitive coupling between the DQD dipole and the detector charge, electrons tunneling in and out of the detector induce time-dependent fluctuations of the DQD energy detuning $\delta$. As a result, the phase difference between the two DQD states picks up a fluctuating component which causes pure dephasing as well as relaxation and excitation, all contributing to the decoherence rate. The detector QD is modelled as a single, spin-degenerate level, coupled to source (S) and drain (D) contacts, via tunnel barriers with rates $\Gamma_\text S$ and $\Gamma_\text D$ respectively. Since $\hbar(\Gamma_\text S+\Gamma_\text D) \gg k_\text B T$, the QD transport is in the lifetime broadened regime. To fully capture the level broadening, we evaluate the detector charge fluctuations within a non-interacting approach, hence neglecting charging effects. Considering the experimentally relevant regime $\Gamma_\text S+\Gamma_\text D,\omega_q \gg \Gamma$, we then arrive at an effective master equation for the DQD density matrix, with a total decoherence rate $\Gamma=\Gamma_\phi+\Gamma_0$. Here $\Gamma_0/2\pi = 710$~MHz is the decoherence rate due to other sources (when the detector is set to Coulomb blockade) and the detector induced rate
\begin{equation}
    \label{eq:detDephase}
    \Gamma_\phi=\frac{4\delta^2}{E^2}\left\{\delta^2 C_\text d(0)+t^2\left[C_\text d(E)+C_\text d(-E)\right]\right\}
\end{equation}
where $C_\text d$ is the frequency dependent dot charge correlation function \cite{SI}. In Eq.~(\ref{eq:detDephase}) the term proportional to $C_\text d(0)$ gives the pure dephasing while the terms proportional to $C_\text d(E)$ and $C_\text d(-E)$ give the decoherence due to relaxation and excitation, respectively. Equation~(\ref{eq:detDephase}) thus provides a compact expression for the total decoherence in terms of the DQD parameters $E,t,\delta$, the capacitively induced energy shift $\Delta E_\text d$ and the detector QD charge correlation function $C_\text{d}(E)$.  

For a quantitative comparison to the experimental results, we first recall that the DQD transition $(0,1) \leftrightarrow (1,0)$ shifts the detector QD energy level by $\Delta E_\mathrm{d}=\SI{5}{\micro eV}$. Likewise, the detector switching between the charge states $0 \leftrightarrow 1$ induces the same energy shift $\Delta \delta=\Delta E_\text d$ in the DQD detuning. Second, by comparing the detector current $I_\text d$ in Fig.~\ref{fig3}(a) with the theoretical predictions, see \cite{SI}, we obtain the tunnel rates $\hbar\Gamma_S=1.5~\mu$eV and $\hbar\Gamma_D=34~\mu$eV. With these values, for the current $I_\mathrm{d} = 270$~pA in point (c), we obtain from Eq.~(\ref{eq:detDephase}) the added detector induced decoherence of $\Gamma_\phi/2\pi = 250$~MHz. This value is roughly half of the measured value $\Gamma_\phi/2\pi = 440$~MHz in Figs.~\ref{fig2}~(c) and (d). 

We see several possible reasons for this difference: Although our detector QD model accounts for level broadening, it does not account for e.g. Coulomb blockade effects or additional charge fluctuations from excited states in the bias window. Moreover, the charge dynamics of the detector QD is taken to be unaffected by the DQD-state, i.e. neglecting DQD-backaction on the detector, which relates to the information content that the detector carries out from the system via the electrical current. Therefore, on a fundamental level, our model does not account for measurement induced decoherence, i.e. the fact that a weak, continuous measurement of the position of the electron, in the L or R dot, suppresses the DQD coherence. We note that we cannot make a similar comparison for the decoherence in points (e), I and II in Fig.~\ref{fig3}(b), which are in a voltage bias regime where additional transport channels through the QD become energetically allowed (see  Fig.~\ref{fig3}(a)).

\textcolor{black}{However, on a qualitative level our model provides important guidance how to reduce the decoherence without suppressing the sensitivity of QD detectors. As is typically the case for QD devices, our detector QD has asymmetric tunnel couplings, $\Gamma_\text S \ll \Gamma_\text D$. Our theory, where the QD charge fluctuations constitute the source of detector induced decoherence, predicts that (see \cite{SI}) the decoherence rate with reversed voltage bias will be a factor  $\sim \Gamma_\text S/\Gamma_\text D$ smaller compared to the present case. The reason for this is that the charge fluctuations are strongly reduced, due to fewer back-and-forth tunnel events between the QD and the leads, for a voltage bias where the detector is operated across the opaque instead of the transparent tunnel barrier. However, the response in detector current $I_\text d$, and hence detector sensitivity, is predicted to be equal under a reversal of the bias voltage. Testing such symmetries in future experiments will provide an interesting avenue towards establishing the optimal detection scheme with minimal decoherence and maximal detector response.}  

In conclusion, we made an experiment probing the coherence of a DQD two-level system under continuous charge sensing. We showed that the effect of the detector is described by an additional decoherence induced to the DQD. The decoherence rate increases roughly proportional to the detector electron tunneling rate ($2I_\mathrm{d}/e$) and is a factor of two larger than a theory estimate based on detector inducted detuning fluctuations. Therefore, this device architecture with tunable detection forms an ideal platform to test various theoretical predictions and quantum measurement concepts~\cite{PhysRevLett.79.3740,PhysRevB.56.15215,PhysRevB.61.2737,PhysRevB.63.125326,oxtoby2006,PhysRevB.81.075301,hatridge2013,hell2014,hell2016,degen2017,bethke2020} experimentally.

We thank K. Ensslin, M. Leijnse and A. Wacker for fruitful discussions, and the Knut and Alice Wallenberg Foundation through the Wallenberg Center for Quantum Technology (WACQT), the European Union (ERC, QPHOTON, project number 101087343), Swedish Research Council (Dnr 2019-04111), the Foundational Questions Institute, a donor advised fund of Silicon Valley Community Foundation (grant number FQXi-IAF19-07), the Swiss National Science Foundation
(Eccellenza Professorial Fellowship PCEFP2\_194268) and NanoLund for funding.

\bibliography{master}

\end{document}


\author{Subhomoy Haldar}
\email{subhomoy.haldar@ftf.lth.se}
\address{NanoLund and Solid State Physics, Lund University, Box 118, 22100 Lund, Sweden}
\author{Morten Munk}
\address{Physics Department and NanoLund, Lund Universityd, Box 118, 22100 Lund, Sweden}
\author{Harald Havir}
\address{NanoLund and Solid State Physics, Lund University, Box 118, 22100 Lund, Sweden}
\author{Waqar Khan}
\address{NanoLund and Solid State Physics, Lund University, Box 118, 22100 Lund, Sweden}
\address{Presently at Low Noise Factory AB, 41263 Göteborg, Sweden}
\author{Sebastian Lehmann}
\address{NanoLund and Solid State Physics, Lund University, Box 118, 22100 Lund, Sweden}
\author{Claes Thelander}
\address{NanoLund and Solid State Physics, Lund University, Box 118, 22100 Lund, Sweden}
\author{Kimberly A. Dick}
\address{NanoLund and Solid State Physics, Lund University, Box 118, 22100 Lund, Sweden}
\address{Center for Analysis and Synthesis, Lund University, Box 124, 22100 Lund, Sweden}
\author{Peter Samuelsson}
\address{Physics Department and NanoLund, Lund Universityd, Box 118, 22100 Lund, Sweden}
\author{Patrick P. Potts}
\address{Department of Physics and Swiss Nanoscience Institute, University of Basel,
Klingelbergstrasse 82, 4056 Basel, Switzerland}
\author{Ville F. Maisi}
\email{ville.maisi@ftf.lth.se}
\address{NanoLund and Solid State Physics, Lund University, Box 118, 22100 Lund, Sweden}
\date{\today}

\title{Supplementary Information: \\Coherence of an Electronic Two-Level System under Continuous Charge Sensing by a Quantum Dot Detector}
\maketitle

In section~\ref{sectionI}, we provide additional measurements of the microwave reflection coefficient and the corresponding theoretical fits. In section~\ref{sectionII}, further information is provided on the theoretical model to explain detector induced additional decoherence of the DQD based two-level system and a fitting of the model to the experiment.

\section{Microwave reflection probability with fits}\label{sectionI}
\begin{figure}[h]
  \centering
{\includegraphics[scale=0.52]{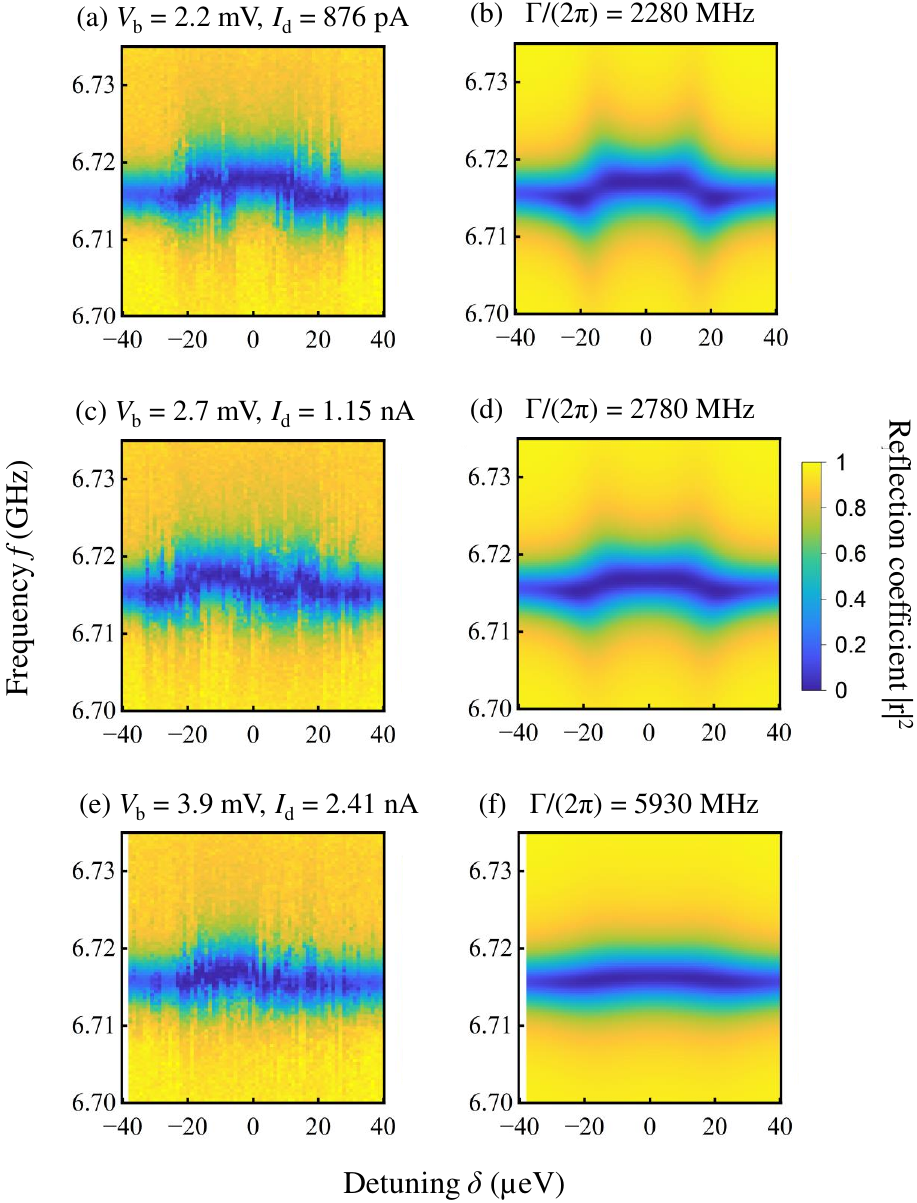}}
  \caption{Reflection coefficient $|r|^2$ as a function of drive frequency and DQD level detuning. Experimental data corresponding to points I,II and the point with largest detector current in Fig. 3b in the main text are shown in panels (a), (c) and (e). The corresponding best fits based on the theory in (1) - (2) in the main text are shown in panels (b), (d) and (f), respectively.}
\label{figapp0}
\end{figure}

In Fig.~3(b) in the main text, six data points for the decoherence rate are shown, obtained by fitting the microwave reflection coefficient data to the theory of Eqs. (1) - (2) in the main text. The measurement data and the corresponding fits giving data points (a), (c) and (e) are given in Fig.~2 in the main text. In Fig.~\ref{figapp0} we present the corresponding information for the data points I, II and the point at largest detector current.
The fitting procedure is described in detail below Eq. (2) in the main text.

\section{Decoherence model and theory}\label{sectionII}
\begin{figure}[b]
  \centering
{\includegraphics[scale=0.85]{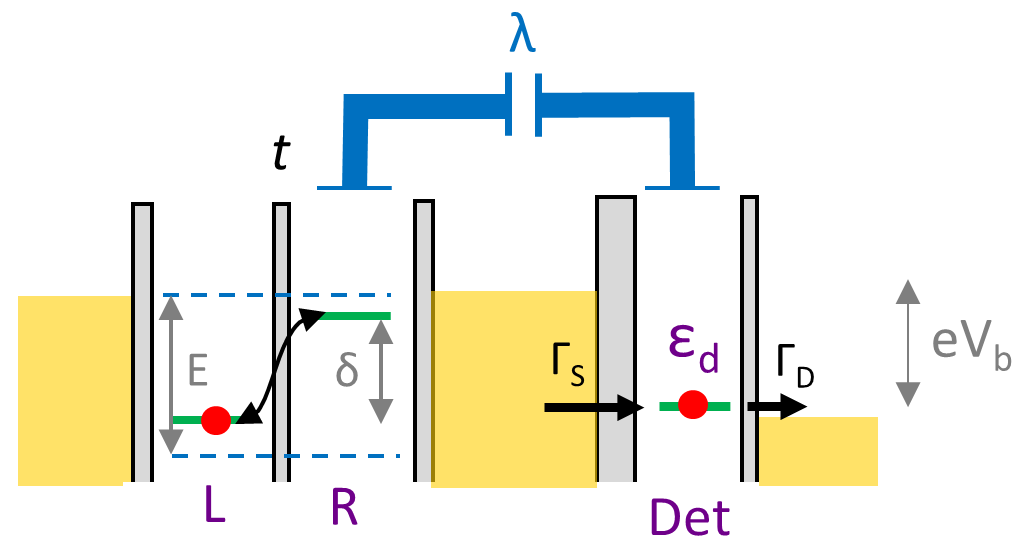}}
  \caption{A schematic of the DQD capacitively coulpled to the detector QD. Level energies, tunnel rates and coupling strengths discussed in the text.}
\label{figapp}
\end{figure}

In this section we will derive a master equation for the reduced density matrix $\hat \rho$ of the DQD, accounting for the coupling to the detector and additional sources of decoherence and relaxation. We here ignore the coupling to the resonator, assuming that we can extend the resulting master equation with terms describing the coupling to the resonator. Moreover, we assume that the effect of the DQD-state on the detector current is small and can be neglected when evaluating the effect of the detector operation on the decoherence in the DQD. Throughout the section, for simplicity we work with $\hbar=1$. A sketch of the DQD-detector system is shown in Fig.~\ref{figapp}

\subsection{Hamiltonian}
We start with a fully unitary description with the Hamiltonian for the joint DQD-detector system
\begin{equation}
    \label{eq:hamiltonian}
    \hat{H}_{\rm tot} = \hat{H}_{\rm DQD} +\hat{Q}\left[r\delta\hat{n}+\lambda \delta\hat{\xi}\right]+\hat{H}_n+\hat{H}_{\rm d}.
\end{equation}
Here the DQD Hamiltonian is
\begin{equation}
    \label{eq:hatdqd}
    \hat{H}_{\rm DQD} = \frac{\delta}{2}\hat{Q}+t\left(|L\rangle\langle R|+|R\rangle\langle L|\right)=\frac{E}{2}\hat{\sigma}_z,
\end{equation}
where $\delta$ is the DQD-detuning, $t$ the tunnel coupling between the two DQD-states $|L \rangle$ and $|R\rangle$ and the effective DQD charge dipole operator
\begin{equation}
\hat{Q} = |R\rangle\langle R|-|L\rangle\langle L|.
\end{equation}
The energy splitting between the hybridized DQD-states (see Fig.~\ref{figapp}) is
\begin{equation}
    \label{eq:omega}
    E = \sqrt{4t^2+\delta^2}
\end{equation}
and $\hat \sigma_z$, as well as other Pauli matrices, is in the basis of hybridized DQD-states.

The second term in the Hamiltonian $\hat{H}_{\rm tot}$ in Eq.~\eqref{eq:hamiltonian} describes the coupling between the effective DQD-dipole and charge fluctuations. Here we consider both the detector occupation fluctuations  $\delta\hat{\xi}=\hat{\xi}-\langle\hat{\xi}\rangle_0$, coupling to the DQD-dipole with a strength $\lambda$, and fluctuations from other nearby charges $\delta\hat{n}=\hat{n}-\langle\hat{n}\rangle_0$, coupled with a strength $r$. The average $\langle\cdot\rangle_0$ denotes the ensemble average with a reference state $\hat{\rho}_{\rm tot} = \hat{\rho}\otimes \hat{\rho}_n\otimes\hat{\rho}_{\rm d}$, where $\hat{\rho}$ describes the DQD, $\hat{\rho}_n$ the nearby charges (usually taken to be a thermal state) and $\hat{\rho}_{\rm d}$ the steady state of the detector. The total detuning of the DQD can thus be written as
\begin{equation}
\delta = \delta_0+2r\langle\hat{n}\rangle_0+2\lambda\langle\hat{\xi}\rangle_0
\end{equation}
where $\delta_0$ is the bare detuning. We note that the difference in the effective detuning between an occupied and empty detector is given by $\Delta \delta = 2\lambda$.

The third term in the Hamiltonian in Eq.~(\ref{eq:hamiltonian}), $\hat{H}_n$, describes the time evolution of the nearby charges and similarly, the fourth term $\hat{H}_{\rm d}$ describes the time-evolution of the detector (including tunneling on and off the detector dot). We assume that there is no tunneling on and off the DQD and that there is a single excess electron residing in the DQD.

\subsection{Master equation}
Starting from the Hamiltonian in Eq.~\eqref{eq:hamiltonian}, we can derive a Markovian master equation \textcolor{black}{for $\hat \rho$} in the secular approximation. For this, we first need the operator $\hat{Q}$ in the interaction picture
\begin{eqnarray}
    \label{eq:qint}
    \tilde{Q}(t)&=&e^{i\hat{H}_{\rm DQD}t}\hat{Q}e^{-i\hat{H}_{\rm DQD}t}  \nonumber \\
&=& \cos(\theta)\hat{\sigma}_z-\sin(\theta)\left(e^{iE t}\hat{\sigma}^\dagger+e^{-iE t}\hat{\sigma}\right),
\end{eqnarray}
with the mixing angle given by  $\cos(\theta)=-\delta/E$ and \textcolor{black}{$\hat \sigma=\hat \sigma_-,\hat \sigma^{\dagger}=\hat \sigma_+$}. Using standard methods, we may derive the master equation
\begin{eqnarray}
    \label{eq:masterdec}
    &\partial_t \hat{\rho}&= -i\frac{E+\Delta}{2}[\hat{\sigma}_z,\hat{\rho}]+(\Gamma_{\rm d,\varphi}+\Gamma_{\rm d})\mathcal{D}[\hat{\sigma}_z]\hat{\rho} \nonumber \\
&+&(\gamma_{\downarrow}+\gamma_{\rm d,\downarrow})\mathcal{D}[\hat{\sigma}]\hat{\rho}+(\gamma_{\uparrow}+\gamma_{\rm d,\uparrow})\mathcal{D}[\hat{\sigma}^\dagger]\hat{\rho},
\end{eqnarray}
where the dissipator $\mathcal{D}[\hat \sigma]\hat \rho=\hat \sigma \hat \rho \hat \sigma^{\dagger}-\{\hat \sigma^{\dagger}\hat \sigma,\hat \rho\}/2$. The Lamb-shift $\Delta$ results in a renormalization of the DQD energy and will be dropped in the following.

The dephasing, relaxation and excitation rates due to the nearby charge fluctuations are
\begin{equation}
    \label{eq:ratesnoise}
    \begin{aligned}
    &\Gamma_\varphi = r^2\cos^2(\theta) \int_{-\infty}^\infty ds C_n(s),\\
  &\gamma_{\downarrow}=r^2\sin^2(\theta)\int_{-\infty}^\infty ds e^{iE s}C_n(s)\hspace{.5cm} \\ 
&\gamma_{\uparrow}=r^2\sin^2(\theta)\int_{-\infty}^\infty ds e^{-iE s}C_n(s),
    \end{aligned}
\end{equation}
with the bath correlation function
\begin{equation}
    \label{eq:bathcorr}
    C_n(s)={\rm Tr}\{e^{i\hat{H}_ns}\delta\hat{n}e^{-i\hat{H}_ns}\delta\hat{n}\hat{\rho}_n\}.
\end{equation}
The total decoherence in the absence of any charge fluctuations in the detector then reads
\begin{equation}
    \Gamma_0 = 4\Gamma_\varphi + \gamma_\downarrow+\gamma_\uparrow.
\end{equation}
In the main text $\Gamma_0$ is determined from the experiment, with the detector tuned away from its operation point.

Along the same lines, the dephasing, relaxation and excitation rates due to the detector operation read
\begin{equation}
    \label{eq:ratesdet}
    \begin{aligned}
    &\Gamma_{\rm d,\varphi} = \lambda^2\cos^2(\theta) \int_{-\infty}^\infty ds C_{\rm d}(s), \\
  &\gamma_{\rm d,\downarrow}=\lambda^2\sin^2(\theta)\int_{-\infty}^\infty ds e^{iE s}C_{\rm d}(s)\hspace{.5cm} \\ 
&\gamma_{\rm d,\uparrow}=\lambda^2\sin^2(\theta)\int_{-\infty}^\infty ds e^{-iE s}C_{\rm d}(s),
\end{aligned}
\end{equation}
with the detector correlation function
\begin{equation}
    \label{eq:detectorcorr}
    C_{\rm d}(s)={\rm Tr}\{e^{i\hat{H}_{\rm d}s}\delta\hat{\xi}e^{-i\hat{H}_{\rm d}s}\delta\hat{\xi}\hat{\rho}_{\rm d}\}.
\end{equation}
The total decoherence induced by the detector then reads
\begin{equation}
    \Gamma_\phi = 4\Gamma_{\rm d,\varphi} + \gamma_{\rm d,\downarrow}+\gamma_{\rm d,\uparrow}.
    \label{totaldetdeph}
\end{equation}
As pointed out in the main text, the total measured decoherence rate is then $\Gamma=\Gamma_0+\Gamma_\phi$.
The assumptions underlying these result are discussed in Section~\ref{secApprox}.

\subsection{Detector charge fluctuations}
To evaluate the rates in Eq.~\eqref{eq:ratesdet}, we write
\begin{equation}
    \label{eq:detectorcorr2}
    \begin{aligned}
    C_{\rm d}(s) &= 2\langle \hat{d}^\dagger(s)\hat{d}(s)\hat{d}^\dagger(0)\hat{d}(0)\rangle \\&- 2\langle \hat{d}^\dagger(s)\hat{d}(s)\rangle\langle\hat{d}^\dagger(0)\hat{d}(0)\rangle,
    \end{aligned}
\end{equation}
where $\hat{d}(t)$ annihilates an electron on the detector dot at time $t$ with $\hat{\xi}(t)=\hat{d}^\dagger(t)\hat{d}(t)$ and the factor of $2$ takes into account spin degeneracy. The detector is operating in the life-time broadened regime, where $\max[\Gamma_{\rm S},\Gamma_{\rm D}]\gg k_{\rm B}T$, which prevents a Markovian description. To make progress, we neglect interactions between electrons, neglecting any Coulomb blockade effects, which allows us to apply Wick's theorem to find
\begin{equation}
    \label{eq:detectorcorr3}
    C_{\rm d}(s) = 2G^<(-s)G^>(s),
\end{equation}
with the lesser and greater Green's functions
\begin{equation}
    \label{eq:lessgreat}
    \begin{aligned}
    &G^<(t) = i\langle \hat{d}^\dagger(-t)\hat{d}(0)\rangle,\\
    &G^>(t) = -i\langle \hat{d}(t)\hat{d}^\dagger(0)\rangle.
    \end{aligned}
\end{equation}
The Fourier transform of the detector correlation function may then be written as
\begin{equation}
    \label{eq:detectorcorromega}
    \begin{aligned}
    C_{\rm d}(\nu) &= 2\int_{-\infty}^\infty ds e^{i\nu s}  G^<(-s)G^>(s) \\&=  \frac{1}{\pi} \int_{-\infty}^\infty d\omega G^<(\omega)G^>(\omega +\nu).
    \end{aligned}
\end{equation}
The Green's function of a single-level quantum dot in frequency space read \cite{jauho1994}
\begin{equation}
    \label{eq:greensf}
    \begin{aligned}
    &G^<(\omega) = i\frac{f_{\rm S}(\omega)\Gamma_{\rm S}+f_{\rm D}(\omega)\Gamma_{\rm D}}{(\omega-\epsilon)^2+(\Gamma_{\rm S}+\Gamma_{\rm D})^2/4},\\
    &G^>(\omega) = i\frac{[1-f_{\rm S}(\omega)]\Gamma_{\rm S}+[1-f_{\rm D}(\omega)]\Gamma_{\rm D}}{(\omega-\epsilon)^2+(\Gamma_{\rm S}+\Gamma_{\rm D})^2/4},
    \end{aligned}
\end{equation}
where we introduced the effective dot level energy $\epsilon=\epsilon_{\rm d} + \alpha_{\rm S} eV_{\rm S}+\alpha_{\rm D}eV_{\rm D}+\alpha_{\rm d}eV_{\rm d}$ where $\epsilon_{\rm d}$ is the bare dot energy, $V_{\rm S},V_{\rm D}$ the applied biases on source and drain contacts, $V_{\rm d}$ the bias on the gate controlling the dot occupation and $\alpha_{\rm S},\alpha_{\rm D}, \alpha_{\rm d}$ describe the relative capacitive couplings between the source, drain, gate and the dot charge ($\alpha_{\rm S}+\alpha_{\rm D}+\alpha_{\rm d}$=1 from gauge invariance). The voltage bias across the detector dot, shown in Fig.~\ref{figapp}, is $V_{\rm b}=V_{\rm S}-V_{\rm D}$

We further introduced the Fermi functions
\begin{equation}
    \label{eq:fermi}
    f_{\rm S/D}(\omega)=\frac{1}{1+e^{(\omega-eV_{\rm S/D})/k_{\rm B}T}},
\end{equation}
where $T$ is the electronic temperature of the contacts.

Inserting Eqs.~\eqref{eq:greensf} into Eq.~\eqref{eq:detectorcorromega} results in
\begin{widetext}
\textbf{\begin{equation}
C_{\rm d}(\nu)=\frac{1}{\pi}\int d\omega \frac{f_{\rm S}(\omega)\Gamma_{\rm S}+f_{\rm D}(\omega)\Gamma_{\rm D}}{(\omega-\epsilon)^2+(\Gamma_{\rm S}+\Gamma_{\rm D})^2/4}\frac{[1-f_{\rm S}(\omega+\nu)]\Gamma_{\rm D}+[1-f_{\rm S}(\omega+\nu)]\Gamma_{\rm D}}{(\omega+\nu-\epsilon)^2+(\Gamma_{\rm S}+\Gamma_{\rm D})^2/4}.
\end{equation}}
\end{widetext}
From Eqs. (\ref{eq:ratesdet}) and (\ref{totaldetdeph}) and the expression for the mixing angle $\theta$ we then get the total decoherence rate from the detector
\begin{eqnarray}
    \Gamma_\phi=\frac{4\lambda^2}{E^2}\{\delta^2 C_\text d(0)+t^2\left[C_\text d(E)+C_\text d(-E)\right]\}.
    \label{dephratefinal}
\end{eqnarray}
Within the same approximations as for the correlation function, the detector current reads \cite{jauho1994}
\begin{eqnarray}
I_{\rm d}=\frac{e}{\pi}\int d\omega \frac{\Gamma_{\rm S}\Gamma_{\rm D}[f_{\rm S}(\omega)-f_{\rm D}(\omega)]}{(\omega-\epsilon)^2+(\Gamma_{\rm S}+\Gamma_{\rm D})^2/4}.
\label{detcurr}
\end{eqnarray}
\subsection{Comparison to experimental results}
\begin{figure}[b]
  \centering
{\includegraphics[scale=0.32]{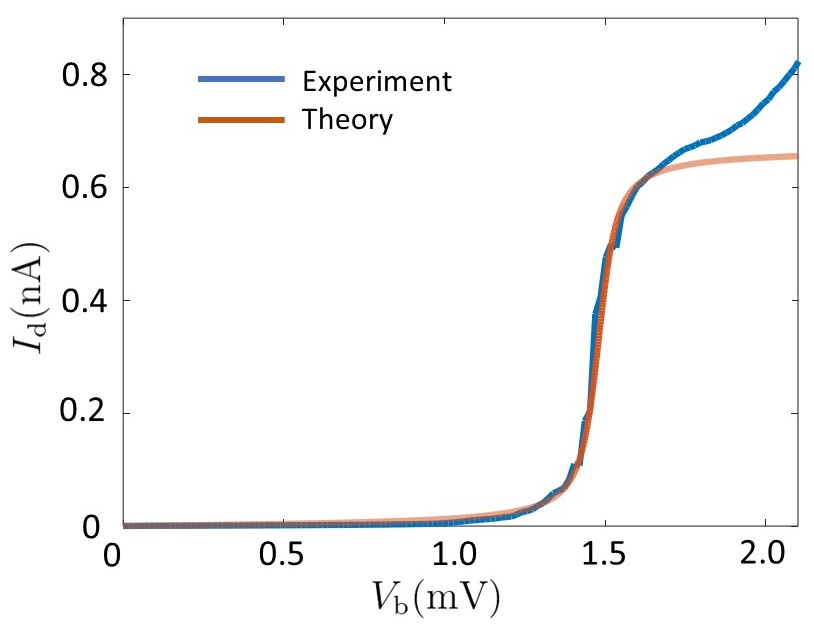}}
  \caption{The measured current of the detector dot, $I_\text d$, as a function of bias voltage $V_\text b$ (blue line), for a dot gate voltage $V_{\text d}=5.050$ V (thin blue line in Fig.~3(a) in the main text). Fit with theory, Eq.~(\ref{detcurr}), see text for details} 
\label{figapp2}
\end{figure}

To compare the theoretical prediction in Eq.~(\ref{dephratefinal}) to the experimentally extracted decoherence rates (Fig.~3(b) in the main text) we proceed in a number of steps. \\
i) By analyzing the slopes of the edges of the Coulomb diamonds \cite{Zhang07,Stampfer08} in Fig.~1(c) in the main text, we find that the capacitive couplings between the detector dot and the S/D contacts are nearly the same, $\alpha_{\rm S} \approx \alpha_{\rm D}$. Moreover, the capacitive coupling of the dot to the gate controlling the dot occupation (see Fig.~1(a) in the main text) is much smaller than the coupling to the S and D contacts, $\alpha_{\rm d}\ll \alpha_{\rm S},\alpha_{\rm D}$. This gives the single capacitive division parameter $\alpha = 0.5$ used in the main text. \\
ii) Next, we compare the detector current prediction in Eq.~\ref{detcurr}) to the measured current in Fig.~2(a) in the main text, a fit is shown in Fig.~\ref{figapp2}. 
The sum of the dot barrier tunnel rates determines the width of the current step in the figure, giving $\Gamma_\text S+\Gamma_\text D=35~\mu $eV. Moreover, the height of the current step is $2e\Gamma_\text S\Gamma_\text D/(\Gamma_\text S+\Gamma_\text D)=0.65$~nA. By further analyzing the differential conductance $dI_\text d/dV_\text b$ at the Coulomb blockade edges (see Fig. 1 (c) in main text), following Refs. \cite{D2NR03459J,PhysRevB.72.233402} it is found that the drain barrier is the most transparent of the two. Together, this gives $\Gamma_\text D=34~\mu$eV and $\Gamma_\text S=1.5~\mu$eV. The bare dot level energy $\epsilon_\text d$ of the model determines the position of the current step in Fig. \ref{figapp2}. To give an overview of the extracted parameters, both in the main text and here, we collect all energies and rates in Tab. \ref{table1}.

\begin{table}[h]
\begin{tabular}{|ccc|}
\hline
  System & Energy  & Angular  \\
 quantity    &  ($\mu$eV) & freq. (GHz) \\ \hline
 $\Gamma_\text S$ &  1.5 & 2.3 \\ 
 $\Gamma_\text D$ & 34 & 52 \\ 
 $k_\text B T$ &  5.2 & 7.9 \\ 
$\lambda=\Delta E_\text d$ & 5.0 & 7.6 \\ 
$E$& 27.8 & 42.2 \\
$t$& 11 & 17 \\
$\delta$ & 16 & 24 \\
$\Gamma_0$ & 2.9 & 4.4 \\ \hline
\end{tabular}
\caption{Extracted values for system energies and rates, for clarity given in both $\mu$eV and GHz.}
\label{table1}
\end{table}

\begin{figure}[t]
  \centering
{\includegraphics[scale=0.32]{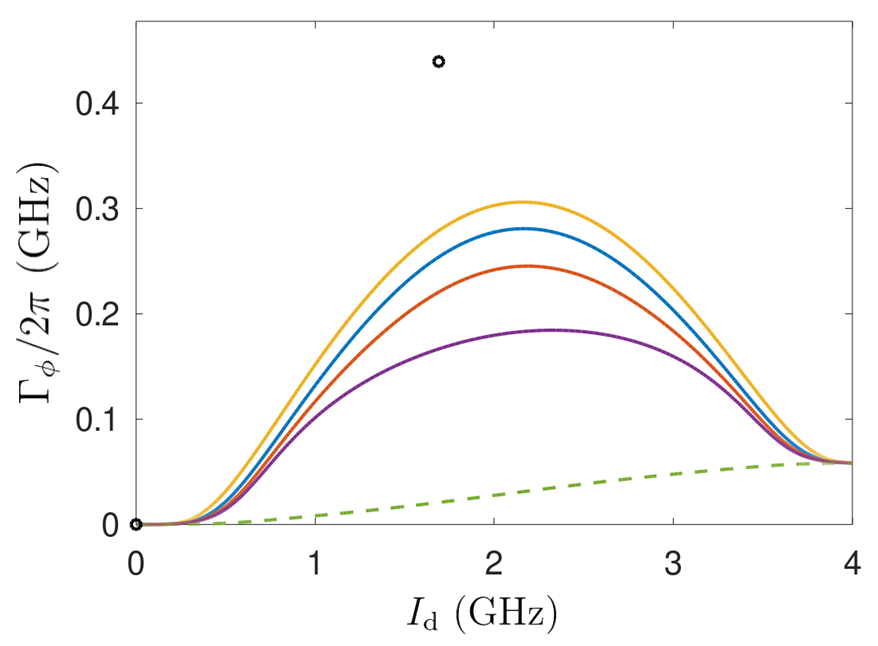}}
  \caption{The decoherence rate due to the detector $\Gamma_\phi$ as a function of detector dot current $I_\text d$, evaluated from Eq.~(\ref{dephratefinal}). The electronic temperature $T=0$ (purple line), $30$ mK (red line), $60$ mK (blue line) and $100$~mK (yellow line). For comparison, the decoherence rate under reversed voltage bias $V_\text b$ is shown with a dashed line. The extracted experimental values  for $I_\text d \approx 0$~nA and $0.27$~nA (denoted points (a) and (c) Fig.~3(b) in the main text) are shown with black circles. For other parameters see text.} 
\label{figapp3}
\end{figure}

We note that since  $\Gamma_\text S+\Gamma_\text D \gg k_\text BT=5.2~\mu$eV, where the estimated electron temperature $T=60$~mK, the current expression in Eq.~(\ref{detcurr}) can be evaluated in the zero temperature limit, giving
\begin{eqnarray}
I_\text b&=&2e\frac{\Gamma_{\rm S}\Gamma_{\rm D}}{\pi(\Gamma_{\rm S}+\Gamma_{\rm D})}\left(\mbox{atan}\left[\frac{2(eV_{\rm D}-\epsilon)}{\Gamma_{\rm S}+\Gamma_{\rm D}}\right] \right. \nonumber \\
 &-&\left. \mbox{atan}\left[\frac{2(eV_{\rm S}-\epsilon)}{\Gamma_{\rm S}+\Gamma_{\rm D}}\right]\right).
\label{curreq}
\end{eqnarray}
We stress that the theory only describes a single level dot and hence can only be expected to capture the experimental curve up to and across the step, that is for $V_\text b \lesssim 1.6$~mV. For these bias voltage values, the transport through the dot in the experiment can be described as, once the Coulomb blockade is lifted, tunneling into and out of a single, effective level, with total source and drain tunnel rates $\Gamma_\text S$ and $\Gamma_\text D$, To describe the observed current steps at $V_\text b>1.7$~mV, in the model additional levels in the dot would have to be included to account for the additional, effective transport channels opening up in the experiment. 
\\
iii) From the experimental characterization of the DQD described in the main text, we have the parameters $\lambda=\Delta E_\text d=5.0~\mu$eV, $t=18~$GHz, $\delta=16~\mu$eV, $E=42.2~$GHz. Together with the values for $\alpha,\epsilon_\text d, \Gamma_\text S$ and $\Gamma_\text D$ deduced from the fit to the measured current (see Tab. \ref{table1}), we can evaluate the theoretically predicted decoherence rate $\Gamma_\phi$ without any additional fitting parameters. Following Fig.~3(b) in the main text, the decoherence rate is plotted in Fig.~\ref{figapp3} as a function of the detector dot current, $I_\text d$, for the different temperatures.
A number of interesting conclusions can be drawn from the figure. \\
$\bullet$  The solid lines in Fig. \ref{figapp3} give the decoherence rate for detector currents $0 < I_\text d <4$~GHz, that is, across the current step in Fig. \ref{figapp2}, at $V_\text b$ around $1.5$~meV. For these bias voltages, the detector QD energy level is close to the Fermi level of the D contact. As pointed out above, the tunnel barrier between the drain and the QD is much more transparent than the barrier between the source and the QD, $\Gamma_\text D \gg \Gamma_\text S$. This leads to larger charge fluctuations and hence more decoherence compared to the situation when the QD energy level is close to the S reservoir Fermi level, shown with a dashed line in Fig. \ref{figapp3}.\\
$\bullet$ In terms of the bias voltage $V_\text b>0$, this implies that the decoherence rate is considerably lowered by reversing the bias, $V_\text b \rightarrow -V_\text b$, equivalent to interchanging the tunnel rates $\Gamma_\text S \longleftrightarrow \Gamma_\text D$. Quantitatively, from Fig. \ref{figapp3} (and similar results for other asymmetric tunnel rates, not shown here), the reduction in decoherence rate can reach values of the order $\sim \Gamma_\text S/\Gamma_\text D$ for the optimal operation point. However, the detector current $I_\text d$, and hence the detection sensitivity, is invariant under a voltage bias reversal. This large suppression of the decoherence rate by operating the detector dot across the most opaque barrier is the key qualitative prediction of our model. \\
$\bullet$ As is also seen in Fig. \ref{figapp3}, the decoherence rate for positive bias, $V_\text b>0$, (solid lines) is predicted to increase monotonically with temperature $T$, while the rate for negative bias,  $V_\text b<0$, (dashed line) is essentially independent on temperature. This is directly related to the observation that the positive and negative bias correspond to detector operation across the transparent and opaque barrier respectively - the temperature has a much stronger effect on the larger charge fluctuations, i.e., for the transparent barrier. \\  
$\bullet$ The calculated decoherence rate at the temperature $T=60~$mK is smaller than the observed rate, roughly $\sim 60 \%$. The theoretical model can thus not account for all the observed decoherence. There are several possible reasons for this, discussed in the main text. \\

\subsection{Approximations}
\label{secApprox}
To derive Eq.~\eqref{eq:masterdec}, we needed to assume that $C_{\rm d}(s)$ decays on a time-scale that is faster than the rate of change of the density matrix (in the interaction picture). This is justified if
\begin{equation}
    \label{eq:ineqmarkov}
    \Gamma_{\rm S}+\Gamma_{\rm D}\gg \Gamma_{\rm d,\varphi},\Gamma_{\varphi},\gamma_{\rm d,\downarrow},\gamma_{\rm d,\uparrow},\gamma_{\downarrow/\uparrow}.
\end{equation}
Since $\Gamma_{\rm L}+\Gamma_{\rm R}=54$~GHz $\gg \Gamma$ (see e.g. Fig.~3(b) in the main text) for the voltage bias regime where the single level model is valid, and $\Gamma=\Gamma_\phi+\Gamma_0=4\Gamma_{\rm d,\varphi} + \gamma_{\rm d,\downarrow}+\gamma_{\rm d,\uparrow}+ 4\Gamma_{\varphi} + \gamma_{\downarrow}+\gamma_{\uparrow}$, the condition in Eq.~(\ref{eq:ineqmarkov}) is satisfied. 

Additionally, we made the secular approximation which is justified if
\begin{equation}
    \label{eq:ineqmarkov2}
    E \gg \Gamma_{\rm d,\varphi},\Gamma_{\varphi},\gamma_{\rm d,\downarrow},\gamma_{\rm d,\uparrow},\gamma_{\downarrow/\uparrow}.
\end{equation}
Again, since $E=42.2$~GHz $\gg \Gamma$, this is ensured.

\bibliography{master}